\documentclass[aps,superscriptaddress,noshowpacs,noshowkeys]{revtex4}
\usepackage{amssymb,amsmath}
\usepackage{color}
\usepackage{graphicx}

\begin{document}
\title{Many faces of  nonequilibrium: \\anomalous transport phenomena in driven periodic systems}
\author{P. H\"anggi}
\affiliation{Institute of Physics, University of Augsburg, D-86135 Augsburg, Germany}
\affiliation{Nanosystems Initiative Munich, Schellingstr. 4, D-80799 M\"unchen, Germany}
\author{J. {\L}uczka}
\affiliation{Institute of Physics,
%and Silesian Center for Education and Interdisciplinary Research,
University of Silesia, 41-500 Chorz{\'o}w, Poland}
\author{J. Spiechowicz}
\affiliation{Institute of Physics,
% and Silesian Center for Education and Interdisciplinary Research,
University of Silesia, 41-500 Chorz{\'o}w, Poland}
%\affiliation{Institute of Physics, University of Augsburg, D-86135 Augsburg, Germany}
%\email{jerzy.luczka@us.edu.pl}
%
\begin{abstract}
We consider a generic system operating under non-equilibrium conditions. Explicitly, we consider an inertial classical Brownian particle dwelling a periodic structure with a spatially  broken reflection symmetry. The particle is coupled to a bath at the temperature $T$ and is driven by an unbiased time-periodic force. In the asymptotic long time regime the particle operates as a Brownian motor exhibiting finite directed transport although no net biasing force acts on the system. Here we review and interpret in further detail recent own research on the peculiar transport behaviour for this setup. The main focus is put on those different emerging Brownian diffusion anomalies. Particularly, within the transient, time-dependent domain the particle is able to exhibit anomalous diffusive motion which eventually  crosses over into normal diffusion only in the asymptotic long-time limit. In the latter limit this normal diffusion coefficient may even show a non-monotonic temperature dependence, meaning that  it is not monotonically increasing with increasing  temperature, but may exhibit instead an extended, intermediate minimum before growing again with increasing temperature.
\end{abstract}
\maketitle

\section{Introduction}

The theory for equilibrium systems is far from being complete. For example, the stationary state of a system strongly interacting with a surrounding thermostat is generally not available  \cite{talkner2020}. Despite this incompleteness for equilibrium setups there are laws in nature which allow us to predict their reaction to an external perturbation. A celebrated example constitutes Le Chatelier-Braun principle \cite{chatelier1884, braun1887a, braun1887b, landau} which loosely speaking states that if a system in equilibrium is subjected to a perturbation, a reaction will occur so that the equilibrium will be shifted towards a new one, counteracting this change. This principle may be regarded as a precursor of a linear response theory \cite{kubo1966, marconi2008} which nowadays is a common tool for predicting the properties of a system in equilibrium perturbed by a external stimuli. An archetypal example is Sutherland-Einstein relation \cite{sutherland1905, einstein1905,chaos2005}, saying that for such setups  the diffusion coefficient is an increasing function of temperature.

Despite many years of active research our current understanding of nonequilibrium physics fundamentals is still incomplete, undoubtedly far beyond what we known for equilibrium systems. Yet much progress has been achieved over the last decades in modelling certain aspects of such systems like stochastic resonance \cite{gammaitoni1998}, noise assisted transport far from equilibrium \cite{kula,hanggi2009}, absolute negative mobility \cite{eichhorn2002a, machura2007, nagel2008, jsm,spiechowicz2014pre, slapik2019}, anomalous diffusion \cite{metzler2014,rysiek,spiechowicz2019njp} or  various recent fluctuation theorems \cite{jarzynski2011, campisi2011}, to name only a few. Here, we aim to demonstrate that nonequlibrium conditions allow for a rich complexity which is not present in a system at thermal equilibrium. The reason behind it is that equilibrium is ruled by various Thermodynamic Laws and symmetries such as for example  {\it detailed balance}, which generally loose their validity if taken out of equilibrium. For this purpose we survey recent research on peculiar transport behaviour occurring in temporally driven periodic system with the particular emphasis put on diffusion anomalies. We rely on the Langevin equation description for nonlinear Brownian motion which, as we shall demonstrate, can be successfully applied also to nonequilibrium systems. It can be derived from a corresponding microscopic Hamiltonian description complemented by fundamentals of equilibrium statistical physics imposed on the thermostat \cite{kubo1966}. The system of interest may look simple at first glance; however, the emerging underlying inertial dynamics is exceptionally rich upon observing that the driven Brownian motion is governed by several parameters which in turn yield a complex dynamics.

\section{Generic model of an inertial Brownian motor}
In this work we consider a classical Brownian motor \cite{hanggi2009}. It is typically modeled as an inertial particle of mass $M$ which moves in a  spatially periodic  potential $U(x)$ which breaks reflection symmetry and additionally  is driven by an unbiased time-periodic force $A\cos{(\Omega t)}$ of amplitude $A$ and angular frequency $\Omega$. The system is coupled to thermostat of  temperature $T$. The corresponding Langevin equation reads
\begin{equation} \label{LL}
	M\ddot{x} + \Gamma\dot{x} = -U'(x) + A\cos{(\Omega t)} + \sqrt{2\Gamma k_B T}\,\xi(t),
\end{equation}
where the dot and the prime denote differentiation with respect to time $t$ and the Brownian particle coordinate $x$, respectively.
The parameter $\Gamma$ stands for the kinetic friction coefficient and $k_B$ denotes the Boltzmann constant. The interaction with thermostat is modeled by $\delta$-correlated, Gaussian white noise $\xi(t)$ of vanishing mean and unit intensity, i.e.,
\begin{equation}
	\langle \xi(t) \rangle = 0, \quad \langle \xi(t)\xi(s) \rangle = \delta(t-s).
\end{equation}
The spatially periodic potential $U(x)$ is assumed to be reflection non-symmetric, i.e. of a ratchet-type \cite{hanggi2009,denisov2014}. As an example, we choose a double-sine form of period $2\pi L$ and  barrier height $2 \Delta U$; explicitly
\begin{equation}
	\label{pot}
	U(x) = -\Delta U\left[ \sin{\left(\frac{x}{L}\right)} + \frac{1}{4}\sin{\left( 2 \frac{x}{L} + \varphi - \frac{\pi}{2}\right)}\right].
\end{equation}

Before we start the analysis of this setup we need to transform the above equation of motion in its dimensionless form. Towards this aim, we introduce a dimensionless distance and time variables for the system under consideration \cite{spiechowicz2015pre, slapik2018}; i.e. we set
\begin{equation}
	\label{scales}
	\hat{x} = \frac{x}{L}, \quad \hat{t} = \frac{t}{\varkappa_0}, \quad \varkappa_0 = \frac{\Gamma L^2}{\Delta U},
\end{equation}
so that the dimensionless form of the Langevin dynamics (\ref{LL}) reads
\begin{equation}
	\label{dimlessmodel}
	m\ddot{\hat{x}} + \dot{\hat{x}} = -\hat{U}'(\hat{x}) + a\cos{(\omega \hat{t})} + \sqrt{2Q} \hat{\xi}(\hat{t})\;.
\end{equation}
Here, the dimensionless potential $\hat{U}(\hat{x}) = U(x)/\Delta U = U(L\hat{x})/\Delta U = \hat{U}(\hat{x} + 2\pi)$ possesses the period $2\pi$ and half of the barrier height is $\Delta {\hat U} = 1$. The remaining parameters are scaled as: $m = M/(\Gamma\varkappa_0)$, $a = (L/\Delta U)A$, $\omega = \varkappa_0\Omega$. The rescaled thermal noise reads \mbox{$\hat{\xi}(\hat{t}) = (L/\Delta U)\xi(t) = (L/\Delta U)\xi(\varkappa_0\hat{t})$} and assumes the same statistical properties as $\xi(t)$, namely $\langle \hat{\xi}(\hat{t}) \rangle = 0$ and \mbox{$\langle \hat{\xi}(\hat{t})\hat{\xi}(\hat{s}) \rangle = \delta(\hat{t} - \hat{s})$}. The dimensionless noise intensity $Q = k_BT/\Delta U$ is the ratio of thermal and half of the activation energy the particle needs to overcome the non-rescaled potential barrier. In order to simplify the notation further, we shall omit the $\wedge$-notation in the above equation (\ref{dimlessmodel}).

The above proposed scaling procedure is not unique as one is free to  define other characteristic time scales of the system described by Eq. (\ref{LL}); namely,
\begin{equation}
	\varkappa_0 = \frac{\Gamma L^2}{\Delta U}, \qquad \varkappa_1 = \frac{M}{\Gamma}, \qquad \varkappa_2^2 = \frac{ML^2}{\Delta U}, \qquad \varkappa_3 = \frac{2\pi}{\Omega} \;.
\end{equation}
Note that only three of them are independent because $\varkappa_0 \varkappa_1 = \varkappa_2^2$. Here we use as the unit of time $\varkappa_0$ , see in Eq. (\ref{scales}) above. This corresponds to the characteristic time scale for an overdamped particle to move from the maximum of the potential $U(x)$ to its minimum. It can be extracted from the equation $\Gamma \dot x = -U'(x)$.  The scale $\varkappa_1$ denotes a relaxation time of the velocity $v=\dot x$ of the free Brownian particle (i.e. for the choice $U(x)= A = 0$) which is obtained from the relation $M\ddot x +\Gamma\dot x=0$. Note that here the dimensionless mass emerges as $m=\varkappa_1/\varkappa_0$; i.e. it equals the ratio of these two characteristic time scales. The quantity $\varkappa_2$ is a characteristic time scale for the conservative system (when $\Gamma=A=0$) and follows from the equation $M \ddot x =-U'(x)$. It is related to the period of the linearized particle oscillations within one potential well. The remaining third time scale $\varkappa_3$ is the period of the external time-periodic force. Thermal fluctuations are modeled here approximately as white noise. In real systems, however, it is never strictly zero but physically typically much smaller than the other time scales.\\

The limit $\Gamma  \longrightarrow \infty$, implying that $ m \longrightarrow 0$, presents an overdamped thermal rocking ratchet dynamics, whose adiabatic and alike its non-adiabatic driving  regimes have been thoroughly studied previously in Refs. \cite{epl1994,lnp1996}; in both, its stochastic dynamics at finite temperatures and as well in its deterministic limit  \cite{epl1994,acta2006}. Remarkably,  this overdamped determinitic regime is already rather complex, exhibiting for example  locking regimes which follow a devil's staircase behaviour \cite{lnp1996,acta2006}.\\
	
The potential (\ref{pot}) has originally been derived for the asymmetric superconducting quantum interference device (SQUID) which is composed of a loop with three capacitively and resistively shunted Josephson junctions
\cite{kautz, zapata1996, spiechowicz2014prb, sterck2005, spiechowicz2015chaos}. The particle coordinate $x$ and velocity $v$ corresponds to the Josephson phase and the voltage drop across the device, respectively. The particle mass stands for the capacitance of the SQUID, the friction coefficient translates to the reciprocal of the SQUID resistance. The time-periodic force corresponds to the modulated external current. The asymmetry parameter $\varphi$ of the potential (\ref{pot}) can be controlled by an external magnetic flux which pierces across the device.

From a mathematical point of view Eq. (\ref{dimlessmodel}) is a second order differential equation additionally complemented by a random force. At first glance it seems simple for undertaking a study. However, note that even the phase space of the noiseless autonomous system modeled by Eq. (\ref{dimlessmodel}) is already three-dimensional $\{x,y=\dot{x},z=\omega t\}$; therefore being minimal for it to display a chaotic dynamics \cite{strogatz}. Moreover, the underlying parameter space $\{m, a, \omega, Q, \varphi\}$ is five-dimensional implying a rich and correspondingly highly complex behaviour.
%Its numerical analysis is by no means straightforward.
The probability density $P(x, v, t)$ for the particle coordinate $x$ and its velocity $v$ obeys a Fokker-Planck equation corresponding to the Langevin equation (\ref{dimlessmodel}) \cite{risken}. It is a parabolic partial differential equations with a time-periodic drift coefficient in phase space of position and velocity. Combining it with a nonlinear periodic potential $U(x)$ together with a five-dimensional parameter space, corresponding analytic time-dependent solutions become in practice unattainable and we are thus  forced to use advanced numerical resources. Details of the latter are elaborated in Ref. \cite{spiechowicz2015cpc}. However, for large dimensionless times $(t \gg 1$) the probability density $P(x, v, t)$ approaches the asymptotic periodic probability distribution $P_{as}(x, v, t)= P_{as}(x, v, t  + \mathsf{T})$ with the periodicity $\mathsf{T}=2\pi/\omega$ of the time-periodic driving $a\cos(\omega t)$ \cite{jung1990,jung1991, jung1993}.
\begin{figure}[t]
	\centering
	\includegraphics[width=0.45\linewidth]{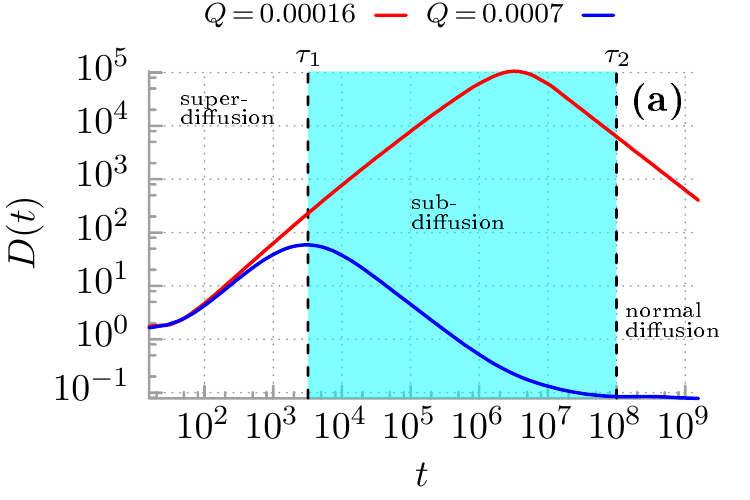}
	\includegraphics[width=0.45\linewidth]{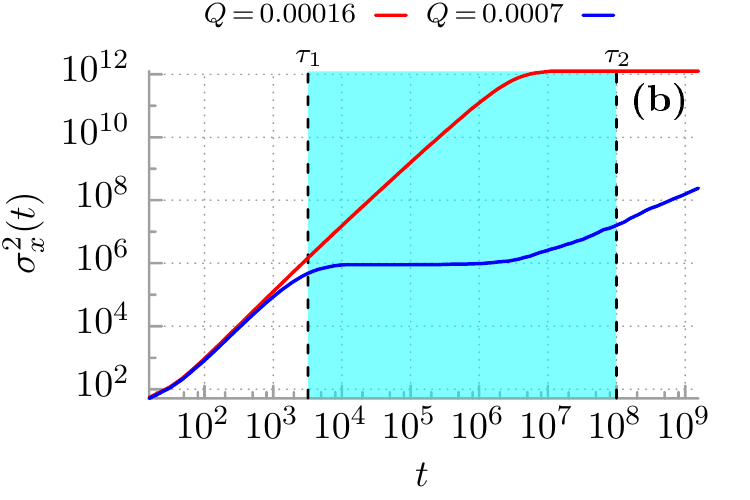}\\
	\includegraphics[width=0.45\linewidth]{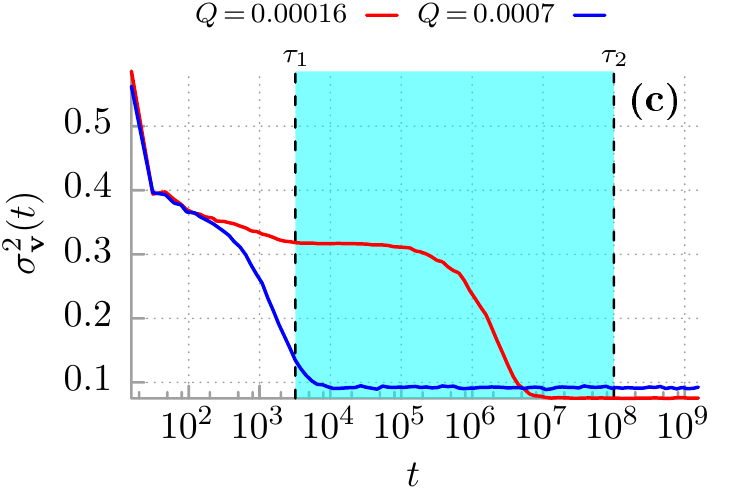}
	\caption{Transient anomalous diffusion of an inertial Brownian particle moving in a periodic potential and driven by a unbiased time-periodic force.   In panel (a) we present the diffusion coefficient $D(t)$ as defined in Eq. (\ref{diff}).  Panel (b) depicts  time evolution of the coordinate variance  $\sigma_x^2(t)$. In panel (c) the period averaged velocity variance  $\sigma_\mathbf{v}^2(t)$ is shown.
%	In panel(c) we present the variance  of the period averaged velocity $\sigma_\mathbf{v}^2(t)$.
Two cases of thermal noise intensity $Q$  proportional temperature are presented (red and blue line).  The region corresponding to the subdiffusive behaviour is for the intensity set at $Q=0.0007$ (region in cyan colour). The remaining parameters are chosen as: $m = 6$, $a = 1.899$, $\omega = 0.403$. The rescaled potential is $U(x)=-\sin(x) -(1/4) \sin(2x)$, which corresponds to $\varphi = \pi/2$. These panels are reproduced from Ref. \cite{spiechowicz2017scirep}.}
	\label{fig1}
\end{figure}

\section{Transient regime: anomalous diffusion}
The diffusion behaviour of the particle dynamics and the spread of its trajectories is conventionally characterized  by the  mean-square deviation (variance) of the particle position $x(t)$ \cite{metzler2014}, namely,
\begin{equation}
	\label{msd}
	\sigma_x^2(t) =  \langle \left[x(t) - \langle x(t) \rangle \right]^2 \rangle = \langle x^2(t) \rangle - \langle x(t) \rangle^2,
\end{equation}
where the averaging $\langle \cdot \rangle$ is over all realizations of thermal fluctuations as well as over the initial conditions for the position $x(0)$ and the velocity $\dot{x}(0)$. The latter is necessary because in the deterministic limit of vanishing thermal noise intensity $Q \to 0$ the dynamics may possess several coexisting attractors thus being non-ergodic and implying that the corresponding results may be affected by a specific choice of those selected initial conditions \cite{spiechowicz2016scirep}. If the coordinate variance grows linearly in evolving time; i.e.,
\begin{equation}
	\label{normal}
	\sigma_x^2(t) =  2Dt
\end{equation}
we refer to  diffusion  as {\it normal}  and the parameter $D$ is termed the diffusion coefficient. Any deviation from this strict linearity qualifies as a process exhibiting anomalous diffusion \cite{hofling2013, metzler2014, meroz2015, zaburdaev2015}. For anomalous diffusion the variance assumes an increasing function of elapsing time, growing either according to a sub-diffusive or a superdiffusive power law \cite{metzler2014}
\begin{equation}
	\label{alpha}
	\sigma_x^2(t)  \sim t^{\alpha}
\end{equation}
Normal diffusion is observed for $\alpha = 1$. The case $0 < \alpha < 1$ refers to  subdiffusion while the case  $\alpha > 1$ is classified as superdiffusion.  It becomes appropriate for the following discussion to consider a time-dependent "diffusion coefficient"  $D(t)$, defined by the relation \cite{spiechowicz2016scirep}
\begin{equation}
	\label{diff}
	D(t) := \frac{\sigma_x^2(t)}{2t}.
\end{equation}
If the behaviour is as in  (\ref{alpha})  then $D(t) \sim t^{\alpha -1}$ and
\begin{itemize}
\item $D(t)$ is time-decreasing for subdiffusion,
\item $D(t)$ is constant for normal diffusion,
\item $D(t)$ is time-increasing for superdiffusion.
\end{itemize}
We stress that only in the asymptotic long time regime with the exponent $\alpha$ approaching unity we find a properly defined, finite diffusion coefficient $D$, i.e.,
\begin{equation}
	D = \lim_{t \to \infty} D(t) < \infty.
\end{equation}
If the diffusion process is anomalous then $D(t)$ either converges to zero (for subdiffusion) or diverges to infinity (for superdiffusion) when $t\to\infty$.
\begin{figure}[t]
	\centering
	\includegraphics[width=0.45\linewidth]{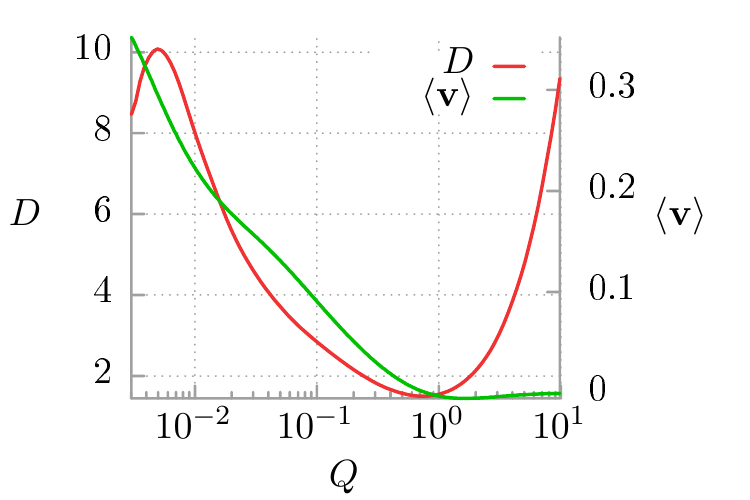}
	\caption{The dependence of the asymptotic diffusion coefficient $D$ (left side ordinate) and the directed velocity
$\langle \mathbf{v} \rangle$  (right side ordinate, see eq. (16))  {\it vs.} the noise intensity $Q$, the latter being proportional to the temperature $T$ of the bath. The chosen parameters are: $m = 6$, $a = 1.899$, $\omega = 0.403$, $\varphi = \pi/2$. The panel reproduced from Ref. \cite{spiechowicz2017chaos}.}
	\label{fig2}
\end{figure}

In panels (a) and (b) of Fig. 1 we depict time evolution of the diffusion coefficient $D(t)$ and the coordinate mean square deviation $\sigma_x^2(t)$, respectively, for two  values of the noise intensity $Q \propto T$. At first glance, it is difficult to identify whether in fact anomalous diffusion takes place by just inspecting the behavior for  $\sigma_x^2(t)$. In distinct contrast, from inspecting instead the behaviour for  $D(t)$ it becomes more facile to differentiate between the two anomalous types of diffusion: superdiffusion occurs in the interval where $D(t)$ increases while the case of decreasing $D(t)$ corresponds to subdiffusion. For an invariant $D(t)$ normal diffusion takes place. In panel (a), the evolution of $D(t)$ can be divided into the three following time-intervals: an early behaviour of superdiffusion $(0, \tau_1)$, an intermediate temporal interval $(\tau_1, \tau_2)$ where subdiffusion emerges over several decades and an asymptotic long time regime $t> \tau_2$ where normal diffusion occurs. The crossover times $\tau_1$ and $\tau_2$ separating these domains can be controlled by the temperature or noise intensity. For a lower temperature (i.e. $Q=0.00016$, see the red curve in panel (a)) the lifetime of superdiffusion is extremely long.  In fact it tends to infinity when $Q\to 0$ (the deterministic case). For $Q = 0.00016$ the superdiffusion regime extends to  $\tau_1 \approx 3.2 \cdot 10^6$. It is difficult to numerically determine $\tau_2$ due to limited stability of the utilized algorithm,  leading to uncontrolled propagation of roundoff- and truncation-errors. However, if we adopt an  extrapolation from other cases then  the time $\tau_2$  is at least of order $10^{11} \sim 10^{13}$. For higher temperature ($Q=0.0007$, the blue curve in panel (a)) the lifetime $\tau_1$ is shorter and it tends to zero when the temperature tends to infinity. For $Q = 0.0007$ the superdiffusion lifetime is  at $\tau_1 \approx 3.2 \cdot 10^3$ and for subdiffusion it is at $\tau_2 \approx 10^8$.  For higher temperatures the  dynamics is initially superdiffusive and approaches a normal diffusion behavior without exhibiting an intermediate subdiffusion time-interval. It is important to note that generally the anomalous diffusion behavior is only of a transient nature and eventually it always tends to normal diffusion in the asymptotic long time limit.

\section{Asymptotic normal long time diffusion: Non-monotonic temperature dependence}
In the standard Sutherland-Einstein relation \cite{sutherland1905, einstein1905} valid for systems at thermal equilibrium the diffusion coefficient $D$ is a monotonically increasing linear function of temperature $T$, i.e.,
\begin{equation}
	\label{einstein}
	D=\mu k_B T,
\end{equation}
where $\mu$ is a mobility coefficient. This is in accordance with our intuition because when temperature grows then thermal fluctuations become larger and in consequence  fluctuations of the particle position also increase. However, for some parameter regimes of our nonequilibrium setup we observe an atypical,  non-monotonic temperature dependence for the emerging asymptotic normal diffusion constant $D$ \cite{spiechowicz2017chaos}. An example is presented with Fig. \ref{fig2}. At low temperatures $Q$ the diffusion coefficient  increases with increasing $Q$ until it reaches a local maximum at $Q\approx 2 \cdot 10^{-5}$ (cf. red line). Then it {\it decreases} towards a minimum at $Q\approx 5\cdot 10^{-3}$ before turning over into a monotonically growing function of $Q$; finally, at sufficiently large values of $Q$, the diffusion coefficient $D$ becomes precisely proportional to $Q$; i.e. to the temperature $T$ of the ambient thermal bath. This high temperature behaviour, however, is not depicted in Fig. \ref{fig1}. The decrease of the diffusion constant with increasing temperature $Q \propto T$ is truly counter-intuitive, being in clear contrast with the Sutherland-Einstein relation (\ref{einstein}) as well as with other known relations such as for example Vogel-Fulcher-like laws \cite{goychuk2014} or an Arrhenius-type behaviour for the diffusion of a Brownian particle in periodic potentials \cite{lifsonjackson,festa1978,htb1990}.

\begin{figure}[t]
	\includegraphics[width=0.45\linewidth]{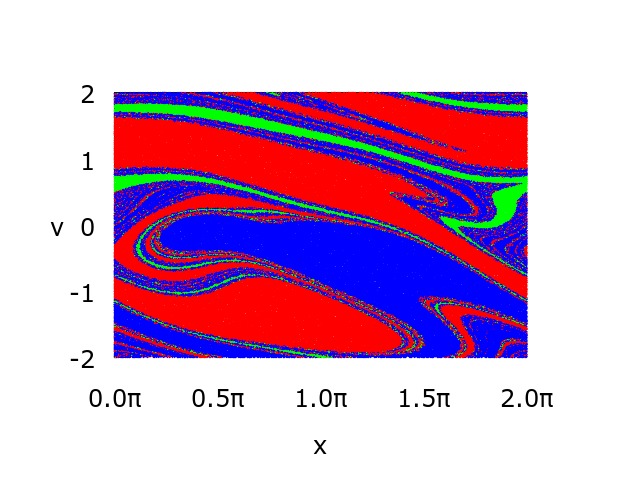}
	\caption{Basins of attraction for the asymptotic long time particle velocity $\mathbf{v}(t)$. The red and blue coloured sets consist of all initial conditions $\{x(0), \dot{x}(0)\}$ eventually evolving to the running states with the positive $v_+  \approx 0.4$ and negative $v_-  \approx -0.4$ velocity, respectively. The green colour marks the set of locked states $v_0  \approx 0$. Parameters are: $m = 6$, $a = 1.899$, $\omega = 0.403$, $\varphi = \pi/2$. For this particular regime the deterministic system (\ref{dimlessmodel}) with $Q = 0$ is non-chaotic. Panel is reproduced from \cite{spiechowicz2016scirep}.}
	\label{fig3}
\end{figure}

\section{Averaged velocity of the Brownian motor}
In order to explain the above two anomalous transport phenomena  one needs first to carefully examine the deterministic structure of the phase space $\{x, v\}$ of all coordinates and velocities of the Brownian motor. For the presented parameter regime, c.f. Figs. 1 and 2, the noiseless system with $Q=0$ is non-chaotic with three coexisting attractors $\{v_+, v_0, v_{-}\}$ in the velocity subspace $\{v\}$. These attractors correspond to running solutions with $v_+ \approx 0.4$ and  $v_{-} \approx -0.4$, and the locked solution  $v_0 \approx 0$. There are three classes of trajectories corresponding to these three states: $x_+(t)\sim 0.4 t$, $x_{-}(t) \sim -0.4 t$ and $x_0(t) \sim 0$.
The basins of attraction for these attractors is shown in Fig. \ref{fig3}. %It is reproduced from our previous paper \cite{spiechowicz2016scirep}.
The red and blue sets consist of all initial conditions $\{x(0), v(0)\}$ evolving into the running states with either positive $v_+ \approx 0.4$ and negative $v_-  \approx -0.4$ velocity, respectively. The green colour regimes  mark the locked states with  $v_0 \approx 0$.
\begin{figure}[t]
	\centering
	\includegraphics[width=0.45\linewidth]{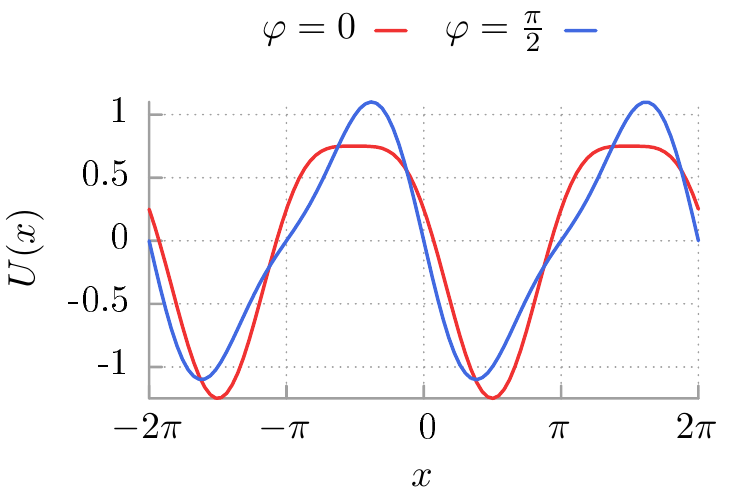}
	\includegraphics[width=0.45\linewidth]{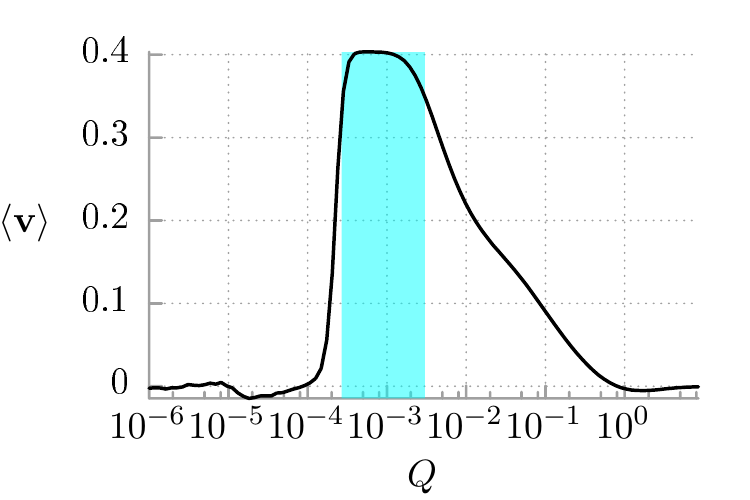}
	\caption{Left panel: The potential given by Eq. (\ref{pot}) depicted in the symmetric case $\varphi = 0$ and its ratchet form for the asymmetry parameter $\varphi =\pi/2$. Right panel: The asymptotic long time averaged directed velocity $\langle \mathbf{v} \rangle$ versus the noise intensity %temperature of the system
$Q$, being proportional to temperature $T$ of the ambient bath. Parameters are: $m = 6$, $a = 1.899$, $\omega = 0.403$, $\varphi = \pi/2$. Right panel is reproduced from \cite{spiechowicz2019chaos}.}
	\label{fig4}
\end{figure}

When the noise intensity is non-vanishing, then thermal fluctuations induce a stochastic dynamics which destabilizes those attractors and leads to random  transitions between its coexisting basins of attraction. This situation is analogous to an escape dynamics from metastable wells in multistable equilibrium systems \cite{htb1990}. Such transitions between the running and/or locked states may generate the transient anomalous diffusion documented with the previous sections. Because we are interested not only in the asymptotic state but also in the full time dynamics it is useful to consider the averaged velocity over the realizations and here additionally also over the temporal driving period of the Brownian motor in presence of thermal noise; i.e.,
\begin{equation}
	\label{dvelocity}
	 \langle \mathbf{v}(t) \rangle = \frac{\omega}{2\pi} \int_t^{t + 2\pi/\omega} ds \, \langle \dot{x}(s) \rangle
\end{equation}
and its variance
\begin{equation}
	\label{vvariance}
\sigma_\mathbf{v}^2(t) = \langle \mathbf{v}^2(t) \rangle - \langle \mathbf{v}(t) \rangle^2 \;.
\end{equation}
%
%where now additional averaging over noise realization should be carried out.
In the asymptotic long time limit these so double-averaged quantities become time-independent, while the solely noise-averaged quantities alone assume a time-periodic function of the asymptotic time-periodic (with period $\mathsf{T}$)  phase-space probability $P_{as}(x, v, t)= P_{as}(x, v, t  + \mathsf{T})$. Put differently, in the asymptotic long time limit, the mean velocity $\langle \dot{x}(t)\rangle$ takes the form of a Fourier series over all possible higher harmonics of the driving force \cite{jung1990,jung1993, gammaitoni1998}.

\begin{equation}
	\lim_{t \gg 1} \,  \langle \dot{x}(t) \rangle = \langle \mathbf{v} \rangle + v_{\omega}(t) + v_{2\omega}(t) + ...
\end{equation}
where $\langle \mathbf{v} \rangle$ is the  time-independent (dc) component while $v_{n\omega}(t)$ denote time-periodic higher harmonic functions of zero average over the  fundamental period $\mathsf{T}=2\pi/\omega$ of the driving. The averaged directed velocity  $\langle \mathbf{v} \rangle$ can also be  obtained from Eq. (\ref{dvelocity}), namely,
\begin{equation}
	\label{v}
	\langle \mathbf{v} \rangle = \lim_{t \gg 1} \, \langle \mathbf{v}(t) \rangle.
\end{equation}
Due to presence of the external driving the Brownian motor is taken far away from thermal equilibrium and a time-dependent nonequilibrium state is reached in the asymptotic long time regime. Since all forces in the right hand side of Eq. (\ref{dimlessmodel}) are non-biased, a necessary condition for the occurrence of directed transport $\langle \mathbf{v} \rangle  \neq 0$ is the breaking of the reflection symmetry of the potential $U(x)$ \cite{hanggi2009, denisov2014}, cf. the left panel of Fig. \ref{fig4}.

In right panel of Fig. \ref{fig4} we plot the directed velocity $\langle \mathbf{v} \rangle$ as a function of temperature $Q \propto T$. For vanishing intensity $Q \to 0$ of thermal fluctuations the directed velocity $\langle \mathbf{v} \rangle \to 0$, which is in agreement with the probability distribution $P(\mathbf{v}(t))$ of the individual asymptotic long time period averaged motor velocity for the deterministic variant of the system (\ref{dimlessmodel}) with $Q = 0$. It is so because of the weighted average over the two running attractors $v_- = 0.4$ and $v_+ = 0.4$ as well as the locked state $v_0 = 0$ yields a vanishing $\langle \mathbf{v} \rangle = 0$. For slightly higher temperature we observe small fluctuations around the deterministic value $\langle \mathbf{v} \rangle = 0$. A further growth of temperature  causes a notable enhancement of the particle velocity $\langle \mathbf{v} \rangle \approx 0.4$. We marked this region with the cyan color. In this regime for the deterministic counterpart of the system there is no directed transport of the motor, however thermal fluctuations induce it. The reason for this behaviour is concealed in thermally activated jumps in the phase space of the nonequilibrium dynamics.

\section{Mechanism responsible for anomalous transport}
Let us now explain the mechanisms which are at the origin for the above presented transport  anomalies: \\

(i) {\it Superdiffusion} \cite{spiechowicz2016scirep}. In the deterministic case $Q = 0$ there are three attractors in the velocity subspace and there are three classes of trajectories associate with them: $x_+(t)\sim 0.4 t$, $x_-(t) \sim -0.4 t$ and $x_0(t) \sim 0$. The overall mean value of the particle position is negligible small $\langle x(t) \rangle \approx 0$, meaning that also $\langle \mathbf{v} \rangle \approx 0$. Moreover, this fact implies that the mean-square deviation $\langle \Delta x^2(t) \rangle = \langle x^2(t) \rangle - \langle x(t) \rangle^2 \approx  \langle x^2(t) \rangle \sim t^2$. As a consequence superdifusive transport takes place which in fact is ballistic diffusion in this deterministic case. This superdiffusive regime is persistent only if $Q \to 0$. For $Q > 0$ thermal noise induces repeated stochastic transitions among the deterministic solutions $x_+(t)$, $x_-(t)$ and $x_0(t)$ which in turn allow for the occurrence of finite directed Brownian motor transport $\langle \mathbf{v} \rangle \ne 0$ if the reflection symmetry of the system is broken, c.f. Fig. \ref{fig4}. In particular, as temperature grows progressively more transitions from the trajectories $x_-(t)$ and $x_0(t)$ to the solution $x_+(t) \sim 0.4 t$ are observed. There occurs even an intensity  interval where almost all particles travel according to $x_+(t) \sim 0.4 t$ as then $\langle \mathbf{v} \rangle \approx 0.4$, c.f. cyan color regime in Fig. \ref{fig4}. The relaxation time of the velocity $\langle \mathbf{v}(t) \rangle$ to its asymptotic long-time value $\langle \mathbf{v} \rangle$ is the same as the lifetime $\tau_1$ for superdiffusion. If the intensity of thermal fluctuations increases the frequency of thermally activated transitions between the deterministic solutions grows greater and therefore the lifetime for superdiffusion decreases.\\

(ii) {\it Subdiffusion} \cite{spiechowicz2017scirep, spiechowicz2019chaos}. For the noisy system $Q \neq 0$ the directed transport velocity $\langle \mathbf{v} \rangle \ne 0$ and the probability for the particle to be in the positive running state  $v_+ \approx 0.4$ grows  whereas the corresponding quantity to stay in the negative running state $v_+ \approx -0.4$ as well as in the locked state $v_0 \approx 0$ decreases. Consequently, the spread of trajectories is smaller and the subdiffusion is developed. It means that once the particles enter the  state $v_+$ they move almost coherently. This argument is supported by the panel (c) of Fig. \ref{fig1} where the velocity fluctuations are significantly reduced within the time interval for subdiffusion. These small, but still finite fluctuations are responsible for the ultraslow subdiffusion where the observed scaling index $\alpha$ in Eq.(\ref{alpha}) is tiny but nonzero $\alpha \ll 1$ \cite{spiechowicz2017scirep}. In the time interval $\tau_1 < t < \tau_2$ the probability for the particle to be in the state $v_+$ is extremely close to unity, meaning that almost all particle trajectories are localized in this regime.
Finally, for sufficiently long times $t > \tau_2$ random dynamics induced by thermal fluctuations again activate jumps between the coexisting trajectories, thus  eventually leading to normal diffusion.\\

(iii) {\it Non-monotonic temperature dependence of the diffusion coefficient} \cite{spiechowicz2017chaos}. Let us focus on two exemplary temperatures $Q_1 < Q_2$, e.g. $Q_1 = 10^{-2}$ and $Q_2 = 1$ in Fig. \ref{fig2}, in order to explain the mechanism responsible for non-monotonic temperature dependence of the diffusion coefficient. Here we observe that $D(Q_1) > D(Q_2)$. For the lower temperature $Q_1 = 10^{-2}$ the averaged directed velocity is $\langle \mathbf{v} \rangle \approx 0.25 > 0$, whereas for larger $Q_2 = 1$ the velocity $\langle \mathbf{v} \rangle \approx 0$, i.e. $\langle \mathbf{v} \rangle (Q_1) > \langle \mathbf{v} \rangle (Q_2) \approx 0$. The latter observation means that for the lower temperature the deterministic structure of the three attractors $\{v_+, v_0, v_- \}$ is still present and plays an important role in controlling the diffusive properties of the system. In particular, because then $\langle \mathbf{v} \rangle > 0$ the majority of trajectories is traveling  with the positive velocity $v_+$, nevertheless however, a still significant fraction of them follows the locked solution $v_0$ and alike also the negative running solution $v_-$. The probability distribution for the particle velocity can approximately  be represented by  a sum of three Gaussians of different mean values, representing the corresponding deterministic solution $\{v_+, v_0, v_- \}$. This causes a large overall spread among the particles and as a consequence also for the corresponding diffusion coefficient. For larger temperature $\langle \mathbf{v} \rangle \approx 0$ the deterministic structure of attractors ceases to be of relevance. In such a case the probability distribution of the particle velocity can be approximated by a single Gaussian with zero mean. Now, the spread between the trajectories is only due to variance of the latter probability distribution which is significantly smaller as compared to the case with lower temperatures. Consequently the diffusion coefficient is reduced. For even larger growing temperature $Q > Q_2$ the diffusion coefficient behaves in a standard way and increases with increasing temperature. This corroborates the finding that for higher temperature the variance of the Gaussian probability distribution for the particle velocity becomes larger.

\section{Summary}
We have shown that for an inertial nonlinear Brownian motor dynamics modeled by an equation which at first glance may appear simple, i.e., for a one-dimensional Newton equation driven by random forces and external driving, the resulting diffusive dynamics manifests an exceptionally rich spectrum of physical phenomena, including  counter-intuitive anomalous transport behaviour. The main reason responsible for these features is that the system operates far away from thermal equilibrium.  Even the asymptotic long-time state is manifest nonequilibrium  and its analytical form is far from being accessible analytically. Therefore, the only applicable scheme for the analysis of this archetype setup is via numerical simulations. The latter can  explain the physics of the occurring phenomena, but only on a qualitative level. A remaining  open question then is to what extent this simple, stylized setup with its already complex behavior can still serve as a trustworthy paradigm for describing those anomalous transport features occurring in more realistic complex systems possessing many more degrees of freedom.

\section*{Acknowledgment}
This work supported by the Grant NCN 2017/26/D/ST2/00543 (J.S. and J.{\L}.) and alike by
the Deutsche Forschungsgemeinschaft (DFG) Grant No. HA1517/42-1 (P.H.).

\end{document}